\documentstyle[aps,epsfig]{revtex}
\def\b{\bar}
\def\d{\partial}

\def\l{\lambda}

\def\m{\mu}
\def\n{\nu}

\def\~{\widetilde}
\def\h{\eta}

\def\bY3{\bar Y_{,3}}
\def\Y3{Y_{,3}}
\def\z{\zeta}
\def\Z{{\b\zeta}}
\def\Y{{\bar Y}}

\def\`{\dot}
\def\be{\begin{equation}}
\def\ee{\end{equation}}
\def\bea{\begin{eqnarray}}
\def\eea{\end{eqnarray}}

\def\fn{\footnote}

\def\mn{{\mu\nu}}

\begin{document}
\twocolumn

\title{Wonderful Consequences of the Kerr Theorem}

\author{Alexander Burinskii\\
Gravity Research Group, NSI Russian
Academy of Sciences\\
B. Tulskaya 52, 115191 Moscow, Russia}
\maketitle

\begin{abstract}
Kerr's multi-particle solution is obtained on the base of the Kerr
theorem. Choosing
generating function of the Kerr theorem $F$ as a product of partial
functions $F_i$ for spinning particles i=1,...k, we obtain a
multi-sheeted, multi-twistorial space-time over $M^4$ possessing
unusual properties. Twistorial structures of the i-th and j-th
particles do not feel each other, forming a type of its internal
space. Gravitation and electromagnetic interaction of the particles
occurs via a singular twistor line which is common for twistorial
structures of interacting particles.
The obtained multi-particle Kerr-Newman solution turns out to be
`dressed' by singular twistor lines linked to surrounding particles.
We conjecture that this
structure of space-time has the relation to a stringy structure of vacuum
and opens a geometrical way to quantum gravity.
\end{abstract}

\bigskip

{\bf Introduction.} It has been mentioned long ago that the
Kerr-Newman solution displays some relationships to the quantum
world. It is the anomalous gyromagnetic ratio $g=2$, as that of the
Dirac electron \cite{Car}, stringy structures \cite{BurStr,BurTwi}
and other features allowing one to construct a semiclassical model
of the extended electron \cite{BurTwi,Isr,Bur0,IvBur,Lop,New1} which
has the Compton size and possesses the wave properties
\cite{BurTwi,Bur0,BurPra}.

One of the mysteries of the Kerr geometry is  the existence of two
sheets of space-time, $(+)$ and $(-)$, on which the dissimilar
gravitation (and electromagnetic) fields are realized, and fields
living on the $(+)$-sheet do not feel the fields of the $(-)$-sheet.
Origin of this twofoldedness lies in the Kerr theorem, generating
function $F$ of which for the Kerr-Newman solution has two roots
which determine two different twistorial structures on {\it the
same} space-time.

In this letter we describe the Kerr's multi-particle solution.
Choosing generating function $F$ of the Kerr theorem  as a product
of partial functions $F_i$ for spinning particles i=1,...k, we
obtain multi-sheeted, multi-twistorial space-time over $M^4$
possessing unusual properties. Twistorial structures of the i-th and
j-th particles do not feel each other, forming a type of its
internal space. Gravitation and electromagnetic interaction of the
particles occurs via a singular twistor line.

This unusual structure of space-time is the direct generalization of
Kerr's twofoldedness and we conjecture that it displays also a
relation to quantum physics.

{\bf The Kerr-Newman metric} can be represented in the Kerr-Schild
form \be g_{\m\n} = \h_{\m\n} + 2 h k_{\m} k_{\n}, \label{ksa} \ee

where $ \h_\mn $ is metric of auxiliary Minkowski space-time
${M}^4$,

\be h=(mr -e^2/2)/(r^2+a^2\cos ^2 \theta),\label{hKS}\ee

and $k_\m (x)$ is a twisting null field, which is tangent to the
Kerr principal null congruence (PNC) which is geodesic and
shear-free \cite{DKS,BurNst}. PNC is determined by the complex
function $Y(x)$ via the one-form \be
 e^3 = du+ \Y d \z  + Y d \Z - Y \Y d v  = P k_\m dx^\m
\label{cong} \ee where $u, \ v, \ \Z , \ \z $ are the null Cartesian
coordinates. Here $P$ is a normalizing factor for  $k_\m$ which
provide $k_0 =1$ in the rest frame.\fn{We replace
the factors $P$ from $e^3$ to function $h$, so $h$ in \cite{DKS} differs
by factor $P^{-2}$.}
The null rays of the Kerr congruence are twistors.
Form of the Kerr PNC is shown on Fig. 1.

{\bf The Kerr theorem} \cite{DKS,Pen} allows one to describe the
Kerr geometry in twistor terms \cite{Pen,BurNst}.

\begin{figure}[ht]
\centerline{\epsfig{figure=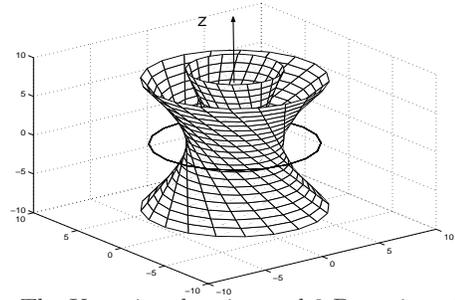,height=4cm,width=6cm}}
\caption{The Kerr singular ring and 3-D section of the Kerr
principal null congruence. Singular ring is a branch line of
space, and PNC propagates from ``negative'' sheet of the Kerr
space to ``positive '' one, covering the space-time twice. }
\end{figure}

It claims that any geodesic and shear-free null congruence in
Minkowski space-time is defined by a function $Y(x)$ which is a
solution of the equation \be F  = 0 , \label{KT}\ee where $F
(Y,\l_1,\l_2)$ is an arbitrary holomorphic function of the
projective twistor coordinates \be Y,\quad \l_1 = \z - Y v, \quad
\l_2 =u + Y \Z .\label{Tw}  \ee

In the Kerr-Schild backgrounds the Kerr theorem acquires a broader
content \cite{BurNst,IvBur1,DKS}, allowing one  to determine the
normalizing function $P$ and complex radial distance $\tilde
r=r+ia\cos\theta ,$

\be P = \d_{\l_1} F - \Y \d_{\l_2} F , \quad \tilde r=PZ^{-1}= -
\quad d F / d Y  \   \label{PF} \ee
which are important
characteristics of the corresponding solutions. The position  of
singular lines, caustics of PNC, corresponds to $\tilde r=0$, and is
determined by the system of equations $ F=0;\quad d F / d Y =0 \ .$

 The
proof of the Kerr theorem in the extended version adapted to the
Kerr-Schild formalism  is given in \cite{BurNst}.

In the original paper \cite{DKS}, the following generating function
$F$ was considered \be F\equiv\phi(Y) +(qY+c)(d\z -Ydv) -(pY+\bar
q)(u+Y\Z) . \label{FKS} \ee

The parameters $ p, q, \bar q, c $ are related to the Killing vector
of the solution $K^\m \d _\m=c\d _u +\bar q \d _\z + q \d _ \Z -p\d
_v ,$  and $ P= pY\bar Y +q Y +\bar q \bar Y +c .$ For the
stationary Kerr solution $p=c=2^{-1/2}, \quad q=\bar q =0$ and
$\tilde r= r+ia \cos\theta$.

It was shown in \cite{IvBur1,KerWil} that function $\phi(Y)$ in
(\ref{FKS}) has to be at most quadratic in $Y$ to provide singular
lines to be confined in a restricted region, which corresponds to
the Kerr PNC up to the Lorentz boosts, orientations of angular
momenta and the shifts of origin.

In the papers \cite{IvBur1,BurNst} another form for this function
was suggested $F=(\lambda _1 -\lambda^0 _1)\check K\lambda _2 -
(\lambda _2 -\lambda^0 _2)\check K\lambda _1$  which is related to
the Newman-initiated \cite{New} complex representation of the Kerr
geometry. In this case, function $F(Y)$ can be expressed via the set
of parameters $q$ which determine the motion and orientation of the
Kerr spinning particle and takes the form $F(Y|q)=A(x|q)Y^2 +B(x|q)Y
+C(x|q)$. The equations (\ref{KT}) can be resolved explicitly,
leading to two roots $Y=Y^\pm (x|q)$ which correspond to two sheets
of the Kerr space-time. The root $Y^+(x)$ determines via
(\ref{cong}) the out-going congruence  on the $(+)$-sheet, while the
root $Y^-(x)$ gives the in-going congruence on the $(-)$-sheet.
Therefore, function $F$ may be represented in the form
$F(Y|q)=A(x|q)(Y-Y^+)(Y-Y^-),$ which allows one to obtain all the
required functions of the Kerr solution in explicit form. The
detailed form of $Y^\pm(x|q)$ is not important for our treatment
here and may be found in \cite{BurNst}.

{\bf Multi-twistorial space-time.} Selecting an isolated i-th
particle with parameters $q_i$, one can obtain the roots $Y_i^\pm
(x)$ of the equation $F_i(Y|q_i)=0$ and express $F_i$ in the form
\be F_i(Y)=A_i(x)(Y-Y_i^+)(Y-Y_i^-).\label{Fi}\ee Then, substituting
the $(+)$ or $(-)$ roots $Y_i^\pm (x)$ in the relation (\ref{cong}),
one determines congruence $k^{(i)}_{\mu}(x)$ and consequently, the
Kerr-Schild ansatz (\ref{ksa}) for metric

\be g^{(i)}_\mn =\eta _\mn + 2h^{(i)} k^{(i)}_{\mu} k^{(i)}_{\nu}
\label{gi},\ee

and finally, the function $h^{(i)}(x)$ may be expressed in terms of
$\tilde r_i= - d_Y F_i ,$  (\ref{PF}), as follows

\be h^{(i)} = \frac m2 (\frac 1 {\tilde r _i} + \frac 1 {\tilde
r_i^*}) + \frac {e^2}{2 |\tilde r_i|^2} \label{hi}. \ee

Electromagnetic field is given by the vector potential

\be  A_\m^{(i)} =\Re e (e/\tilde r _i) k^{(i)}_\m \label{Aisol} .\ee

What happens if we have a system of $k$ particles?
One can form the
function $F$ as a product of the known blocks $F_i(Y)$,

\be F(Y)\equiv \ \prod _{i=1}^k F_i (Y) \label{multi}. \ee

The solution of the equation $F=0$ acquires $2k$ roots $Y_i^\pm$,
and the twistorial space turns out to be multi-sheeted.

\begin{figure}[ht]
\centerline{\epsfig{figure=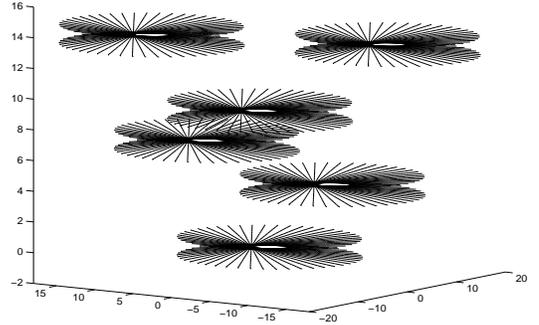,height=5cm,width=7cm}}
\caption{Multi-sheeted twistor space over the auxiliary Minkowski
space-time of the multi-particle Kerr-Schild solution.
Each particle has twofold structure.}
\end{figure}

The twistorial structure on
the i-th $(+)$ or $(-)$ sheet is determined by the equation
$F_i=0$
and does not
depend on the other functions $F_j , \quad j\ne i$. Therefore, the
particle $i$ does not feel the twistorial structures of other
particles. Similar, the condition for singular lines $F=0, \ d_Y
F=0$ acquires the form

\be \prod _{l=1 }^k F_l =0, \qquad   \sum ^k_{i=1} \prod _{l\ne i}^k
F_l d_Y F_i =0 \label{leib} \ee

 and splits into k independent relations

\be F_i=0,\quad \prod
_{l\ne i}^k F_l d_Y F_i =0 \label{kind}. \ee

One sees, that i-th particle does not feel also singular lines of
other particles.  The  space-time splits on the independent
twistorial sheets, and therefore, the twistorial structure related
to the i-th particle plays the role of  its ``internal space''.

It looks wonderful. However, it is a direct generalization of
the well known twofoldedness of the Kerr space-time which remains
one of the mysteries of the Kerr solution for the very long
time.

For spinning particles $|a|>>m$ and the Kerr's black hole horizons
disappear, there appears the old problem of the source of Kerr
solution with the alternative:  either to remove this
twofoldedness or to give it a physical interpretation.
By truncation of the negative sheet, there appears the source
in the form
of relativistically rotating disk \cite{Isr}, bubble \cite{Lop}
or bag \cite{BurBag}.

 Alternative way is to retain the negative sheet, treating it as the sheet
of advanced fields. In this case the source of spinning particle
turns out to be the Kerr singular ring (circular string,
\cite{BurTwi,BurPra}) with the electromagnetic excitations in the
form of traveling waves which generate spin and mass of the particle
(microgeon model  \cite{Bur0,BurPra}).

{\bf Multi-particle Kerr-Schild solution.} Using the Kerr-Schild
formalism with the considered above generating functions $\prod
_{i=1}^k F_i (Y)=0,$ one can obtain the exact asymptotically flat
multi-particle solutions of the Einstein-Maxwell field equations.
Since congruences are independent on the different sheets, the
congruence on the  i-th sheet retains to be geodesic and shear-free,
and one can use the standard Kerr-Schild algorithm of the paper
\cite{DKS}. One could expect that result for the i-th sheet will be
in this case the same as the known solution for isolated particle.
Unexpectedly, there appears a new feature having a very important
consequence.

Formally, we have only to replace $F_i$ by $F=\prod _{i=1}^k F_i
(Y)=\mu _i F_i(Y) ,$ where $\mu _i =\prod _{j\ne i}^k F_j (Y)$ is a
normalizing factor which  takes into account the external particles.
However, in accordance with (\ref{PF}) this factor $\mu _i$ will
appear also in the function $\tilde r  =-d_Y F= -\mu _i d_Y F_i ,$
and in the function $P=\m_i P_i.$

So, we obtain the different result

\be h_i = \frac {m_i(Y)}{2\m_i^3} (\frac
1{\tilde r_i} + \frac 1{\tilde r_i^*})
+ \frac{(e/\mu_i)^2}{2 |\tilde r_i|^2} ,
\label{hKSi} \ee

\be  A_\m^{(i)} =\Re e \frac e {\mu _i \tilde r _i} k^{(i)}_\m
\label{Aren} \ee
which looks like a renormalization of the  mass $m$
and charge $e$.\fn{Function  $m_i(Y)$ is free and
satisfies the condition $(m_i),_\Y =0$. It
  and has to be chosen
in the form $m_i(Y) = m_0\m_i^3 $ to provide reality of metric.}

This fact turns out to be still more intriguing if we note that $\mu
_i$ is not constant, but a function of $Y_i$. We can specify its
form by using the known structure of  blocks $F_i$ \be \mu_i (Y_i)=
\prod _{j\ne i} A_j (x)(Y_i - Y_j^+) (Y_i - Y_j^-) \label{mui}. \ee
The roots $Y_i$ and $Y_j^\pm$ may coincide for some values of $Y_i$,
which selects a common twistor for the sheets $i$ and $j$. Assuming
that we are on the i-th $(+)$-sheet, where congruence is out-going,
this twistor line will also  belong to the in-going $(-)$-sheet of
the particle $j$ . The metric and electromagnetic  field will be
singular along this twistor line, because of the pole $\mu _i \sim
A(x) (Y_i^+ - Y_j^-)$.
Therefore, interaction occurs along a light-like
Schild string which is common for twistorial structures of both particles.
The field structure of this string is similar to the well known structure
of pp-wave solutions.

The equations (\ref{ksa}), (\ref{hKS}) and (\ref{cong}) give the {\it exact
multi-particle solution of the Einstein-Maxwell field equations.}
It follows from the fact
that the equations were fully integrated out in \cite{DKS} and expressed
via functions $P$ and $Z$ before (without) concretization of the form
of congruence, under the only condition that it is geodesic and shear free.
In the same time the Kerr theorem determines the functions
$P$ and $Z$ via generating function $F,$ eq.(\ref{PF}), and the condition of
reality for metric may be provided by a special choice of the free
function $m(Y)$.

The obtained multi-particle solutions show us that, in addition to the usual
Kerr-Newman solution for an isolated spinning particle, there is a series
of the exact `dressed' Kerr-Newman solutions which take into account
surrounding particles and differ by the appearance of singular twistor
strings connecting the selected particle to external particles.
This is a new gravitational phenomena which points out on a probable
stringy (twistorial) texture of vacuum and may open a geometrical way
to quantum gravity.

\begin{figure}[ht]
\centerline{\epsfig{figure=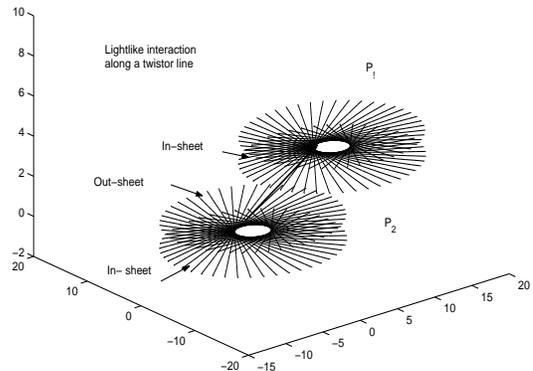,height=6cm,width=7cm}}
\caption{Schematic representation of the lightlike interaction
 via a common twistor line connecting  out-sheet of one particle
 to in-sheet of another.}
\end{figure}

The number of surrounding particles and number of blocks in
the generating function $F$ may be assumed countable. In this case the
multi-sheeted twistorial space-time will possess the properties
of the multi-particle Fock space.

{\bf Acknowledgments.} Author thanks the participants of the seminar
on Quantum Field Theory at the Physical Lebedev Institute for
useful discussion. This work was supported by the RFBR Grant
04-0217015-a and by the ISEP Research Grant by Jack Sarfatti.


\begin{thebibliography}{99}


\bibitem{Car}
B. Carter, Phys.Rev. {\bf 174}, 1559 (1968).
\bibitem{BurStr}
A. Burinskii, Phys.Lett. {\bf A 185} (1994) 441; {\it String-like
Structures in Complex Kerr Geometry.} In: ``Relativity Today'',
Edited by R.P.Kerr and Z.Perj\'es, Akad\'emiai Kiad\'o, Budapest,
1994, p.149.

Phys.Rev. D {\bf 68}, 105004 (2003).

\bibitem{BurTwi} A. Burinskii
Phys.Rev. D {\bf 70}, 086006 (2004).

\bibitem{Isr}
W. Israel, Phys.Rev. D {\bf 2}, 641 (1970).

\bibitem{Bur0}
   A.Ya. Burinskii, Sov. Phys. JETP,  {\bf39}(1974)193.
\bibitem{IvBur} D. Ivanenko and A.Ya. Burinskii,
   Izvestiya Vuzov Fiz. n.5 (1975) 135 (in russian).
\bibitem{IvBur1} D.
   Ivanenko and A.Ya. Burinskii, Izvestiya Vuzov Fiz. n.7 (1978) 113 (in
   russian).

 A.Ya. Burinskii, {\it Strings in the
Kerr-Schild metrics} In: ``Problems of theory of gravitation and
elementary
   particles",{\bf11}(1980), Moscow, Atomizdat, (in russian).

\bibitem{Lop}
C.A. L\'opez, Phys.Rev. {\bf D 30} (1984) 313.

\bibitem{New1} E.T. Newman
Phys.Rev. D {\bf 65}, 104005 (2002).

\bibitem{BurPra}
 A. Burinskii,
{\it Rotating Black Hole, Twistor-String and Spinning Particle},
Czech.J.Phys. {\bf 55}, A261 (2005),
hep-th/0412195, hep-th/0412065

\bibitem{DKS}  G.C. Debney, R.P. Kerr, A.Schild, J. Math.
Phys. {\bf 10}, 1842(1969).
\bibitem{Pen}  R. Penrose, J. Math. Phys. {\bf 8}(1967) 345.

\bibitem{BurNst}
A. Burinskii,
Phys. Rev. D {\bf 67}, 124024 (2003).

 A. Burinskii and R.P. Kerr, {\it Nonstationary Kerr Congruences},
gr-qc/9501012.

A. Burinskii and G. Magli,
Phys.Rev. {\bf D 61}(2000)044017.

\bibitem{KerWil} R.P. Kerr and W.B.  Wilson,
Gen. Relativ. Gravit. {\bf 10}, 273 (1979).

\bibitem{New} E.T. Newman, J. Math. Phys. {\bf 14}, 102 (1973),
R.W. Lind and E.T.Newman, J. Math. Phys. {\bf 15}, 1103 (1974).

\bibitem{BurBag}
A. Burinskii, Grav.\& Cosmology.{\bf 8} (2002) 261.



\bibitem{KraSte} D.Kramer, H.Stephani, E. Herlt, M.MacCallum, ``Exact
Solutions of Einstein's Field Equations'', Cambridge Univ. Press,
1980.

\end{thebibliography}
\end{document}